# Investigating student communities with network analysis of interactions in a physics learning center

Eric Brewe,[1,2] Laird Kramer,[2] and Vashti Sawtelle[2]

[1]*Department of Teaching and Learning, Florida International University, 11200 SW 8th Street, Miami, Florida 33199, USA*
[2]*Department of Physics, Florida International University, 11200 SW 8th Street, Miami, Florida 33199, USA*

Developing a sense of community among students is one of the three pillars of an overall reform effort to increase participation in physics, and the sciences more broadly, at Florida International University. The emergence of a research and learning community, embedded within a course reform effort, has contributed to increased recruitment and retention of physics majors. We utilize social network analysis to quantify interactions in Florida International University's Physics Learning Center (PLC) that support the development of academic and social integration. The tools of social network analysis allow us to visualize and quantify student interactions and characterize the roles of students within a social network. After providing a brief introduction to social network analysis, we use sequential multiple regression modeling to evaluate factors that contribute to participation in the learning community. Results of the sequential multiple regression indicate that the PLC learning community is an equitable environment as we find that gender and ethnicity are not significant predictors of participation in the PLC. We find that providing students space for collaboration provides a vital element in the formation of a supportive learning community.

## I. INTRODUCTION

During the past eight years, the number of physics majors at Florida International University (FIU) has increased by a factor of 400%, far outpacing the national increase in physics. This study explores plausible mechanisms for explaining this increase in number of physics majors by investigating the structure of a learning community which is centered in an informal Physics Learning Center (PLC) at FIU. Many researchers link the academic and social integration of students to increased rates of retention [1–5]. We employ social network analysis (SNA) as both a research framework and a methodology to describe and quantify the existence of a learning community as a plausible explanation of the persistence and retention, and to capture the element of collaboration within the physics department. SNA provides a set of tools for visualization, quantification, and hypothesis testing on networks of actors (in our case students). Thus, SNA provides a way to describe the cultural phenomena that surrounds the Physics Learning Center as a mechanism for understanding the sources of persistence and retention.

Florida International University has established a learning community within the physics department as part of a comprehensive effort to increase participation and success by historically underrepresented students in physics. The comprehensive reform effort rests on three interdependent foundations: (1) reform of several sections of introductory physics by implementing Modeling Instruction, (2) establishing, fostering, and sustaining student collaboration within a physical space known as the Physics Learning Center, and (3) systemic faculty advocacy on behalf of students and institutional reform. Since the inception of The Center for High Energy Physics Research and Education Outreach (CHEPREO) at FIU, the number of physics majors has grown by a factor of 400% when normalized to the growth of FIU. Classroom-based measures alone are insufficient to account for recent increases in enrollment, retention, and persistence seen at FIU since the origin of CHEPREO [6]. Furthermore, exclusively using classroom-based measures inadequately describes cultural phenomena that surround the Physics Learning Center. Characterizing student collaboration in an informal learning environment is complex, as typical classroom-based measures of learning [e.g., grades or Force Concept Inventory (FCI) scores] are not relevant across a wide variety of students who engage in an informal space, and none of these measures capture the relational nature of collaboration. Our primary finding is that the Physics Learning Center at FIU is fostering participation by a variety of students and is inclusive of students of a variety of ethnic background as well as both men and women.

### Physics Learning Center

At FIU, the CHEPREO project initiated the building and sustaining of a classroom to implement reformed instruction, as well as a 12-person conference room and a small student lounge. This complex became known as the Physics Learning Center. The classroom is a 1200 square foot space with five 9 ft × 5 ft student tables making it comfortable for 30 students. The front and back walls are permanently mounted with whiteboards, and typically 30–50 small portable whiteboards are stored at various locations around the room. The classroom has four projectors which are controlled by an instructor station at the front of the class. Two storage closets house data acquisition hardware, including a full range of detectors and 12 laptop computers. The lounge, adjacent to the classroom, has a small kitchen area (refrigerator, sink, toaster oven, coffee pots), a couch, several chairs, and several bookcases. The conference room includes a large table with 12 chairs and one wall with a permanently mounted whiteboard. Several nearby offices house people affiliated with the CHEPREO project and the physics education research group including the CHEPREO coordinator, the teacher-in-residence, and physics education research and high-energy physics graduate students. The classroom is specifically designed to house Modeling Instruction courses. Currently, the classroom is in use approximately 30 hr per week for scheduled classes, but early in the evolution of the PLC the classroom was in use as little as 12 hr per week.

During the early years of the PLC, several of the students in the Modeling Instruction classes expressed interest in having access to the classroom while the room was not in use for classes. A program of "open labs" was developed to grant students access to the room and resources in the classroom. A select set of students were chosen to be CHEPREO Fellows. These Fellows were given open access (keys to the room and computer closets) and responsibility for overseeing the use of the room during open labs. When CHEPREO Fellows were in the room, students had access to the portable whiteboards to work on homework outside of scheduled classes and to use CHEPREO-owned computers.

Student response to the open labs has been overwhelmingly positive, and students can regularly be found in the PLC engaged in a wide variety of activities. Formal activities include meetings of the Society of Physics Students and open labs. However, informal activities are more prevalent. These activities include the myriad things students are involved in, such as studying, eating, planning social events, and generally hanging out. Lower division students also utilize the room to get physics help from upper division students, so students interact across grade levels. Students often can be found collaborating on homework, in physics as well as other subjects such as math, engineering, chemistry, or biology, so the PLC has an interdisciplinary flavor to it. Often faculty visit the PLC, making it a place for students to interact informally with faculty. Nonacademic interactions take place as well, including informal peer advising and social interactions. In short, the PLC serves as the hub for students to become integrated into the social and academic fabric of the university by participating in a learning community.

### 1. Multiple entry points to the PLC learning community

*Modeling Instruction.*—The primary entry point to the learning community in the PLC is through the Modeling Instruction courses that are offered in the PLC. We use the term learning community, as described by Bielaczyc and Collins [7], as consisting of a group of people with varying levels of expertise and a group goal of expanding the community's knowledge. Modeling Instruction courses at FIU operate in a collaborative learning environment, with 30 students in a studio-format class with integrated lab and lecture. Students work together to engage in the process of building, validating, and deploying models [8,9] by engaging in inquiry labs and activities focused on conceptual reasoning and problem solving [8]. Throughout the course students work in small groups of three and also participate in whole class discussions. For example, students may first carry out a lab in groups of three and then use a small portable whiteboard to summarize the findings from their lab, identifying patterns seen in the lab and supporting these patterns with evidence from the lab. Subsequently, the small groups meet in a circle with the whole class in the center of the classroom to compile their findings. The collaborative nature of the Modeling Instruction class includes close cooperation in the small groups as well as consistent interaction with the entire set of members of the class. The interactions inherent in the Modeling Instruction classes, we believe, establish an entry point to the community of learners in the PLC because students see their classmates as valuable resources for learning physics and they are accustomed to the practices of working on whiteboards in the Modeling Instruction classes.

Multiple measures indicate success in the Modeling class, including improved conceptual understanding, attitudes, and odds of success [10,11]. Each of these measures can be viably explained by classroom-based analyses such as the pedagogy or curriculum. Students from Modeling Instruction classes at FIU also have 6.7 times greater odds of success, where success was identified as earning a passing grade and not dropping or withdrawing [10]. In addition, this measure of retention within the introductory course was found to be equitably realized by students that have traditionally been underrepresented in physics (women, Hispanic, and Black students). Beyond the Modeling Instruction class, the number of physics majors has grown by 800% since the early 1990s, a time period during which the university grew by 180%. While measures of learning and attitude can be explained easily by considering the classroom, issues of retention and ultimately persistence merit further consideration of external variables.

*Other pathways.*—The PLC houses a diverse community of learners. Regular users of the PLC include students from several majors; upper and lower division students; men and women; a mixture of ethnic backgrounds; and many that did not take Modeling Instruction for introductory physics. Therefore, if Modeling Instruction is a primary pathway into the PLC, it is certainly not the only pathway. Several other paths lead to participation in the PLC. First, several student organizations, including the Society of Physics Students and the Astronomy club, hold meetings in the PLC. Both clubs are very active, which leads to other opportunities for engaging in activities in the PLC. Word of mouth represents a second pathway. Modern Physics includes students from Modeling Instruction classes and from traditional classes, so when study groups form, students from the Modeling Instruction courses recommend meeting and studying in the PLC. Additionally, FIU has a Learning Assistant (LA) program which draws heavily from Modeling courses while integrating other students. The LA program utilizes the PLC for activities and class meetings, creating another pathway for students to participate. Finally, active advocacy by faculty constitutes another pathway, whereby students are made aware of the PLC and the interactions that take place within the PLC through conversation with faculty members. The multitude of pathways outside of the Modeling Instruction classes represent a complex web of interactions and opportunities for students to learn about the PLC and ultimately to engage in the mix of academic and social activities that take place therein.

### 2. Proposing that the PLC contributes to retention

We consider the role that engagement in a learning community plays in retaining students as a way to understand both the increased success in the Modeling Instruction classes and the increase of number of physics majors at FIU. First, we consider what it means to retain students. We recognize that retention in the introductory course is not synonymous with persistence throughout physics. The American Council on Education finds that incoming Hispanic and Black students who declare a Science, Technology, Engineering and Mathematics (STEM) major typically are still STEM majors after a year, but that Hispanic and Black students switch to non-STEM majors later in their academic careers [12]. This indicates that in order to make inroads on persistence through physics, particularly with the majority Hispanic student population at FIU, consistent support beyond the introductory classroom is critical. Tinto [13] identified precollege characteristics as important factors in student success, but also pointed to students' academic and social integration into the institution as vital to student retention [3]. Substantiating Tinto's findings, Finn and Rock [1] found positive correlations between activities that increase student commitment to their high school and success. Additionally, Kraemer [2] identified informal student-faculty interactions as particularly important for integrating Hispanic students into the community of the university. Tinto [4] argued that the classroom is for many students the only place to meet other students and faculty and as such special attention should be paid to the role of the classroom. Cabrera et al. [14] identified that classrooms involving collaborative learning have significant impacts on increasing diversity among college students. Summarily, finding opportunities to integrate students into the university has been shown to have positive impacts on student retention and persistence. Yet, effective implementations of these findings are rare [5].

Given the research into student retention and persistence, we aver that the Physics Learning Center has contributed vitally to the retention and persistence of physics students at FIU. The Physics Learning Center enables students to engage in many of the ways identified in other research (informal faculty-student interaction, collaborative learning environments, and academic and social integration). Functionally, the PLC serves as a hub for development of a learning community in physics. Further, since student participation in a learning community is linked to increased retention and persistence [4], participation in the PLC should also be associated with persistence in physics.

## II. THEORETICAL FRAMEWORK: A PARTICIPATIONIST PERSPECTIVE ON LEARNING

Acknowledging that student formation of and participation in a learning community are important to retention and persistence is consistent with participationist frameworks on learning. Participationist views on learning can be summarized by suggesting that learning is an ongoing process of transformation of participation, "Learning then occurs as people participate in the sociocultural activities of their community, transforming their understanding, roles and responsibilities as they participate," [15], p 390. Rogoff et al. [15] eloquently argue that learning happens in all instructional settings, and that in all settings participation transforms, but that the way participation transforms is not uniform. For example, in a lecture setting, students learn behavioral and epistemological norms that learning happens through a passive receiving of information. While in active learning environments, students may learn that knowledge is constructed and that students actively participate in this construction. The view of Rogoff et al. helps to clarify the role that participation in a community plays in the learning process [16]. Actively participating in learning within the informal community of the Physics Learning Center provides students with the opportunity to become members of a group whose primary identity is that of a group of physics learners. As a result, we propose that investigating student participation provides a lens through which to view the learning community. In this analysis we target the informal learning community established in the Physics Learning Center and use collaboration as a proxy for participation. To be certain, we do not intend to say that students who occupy a more central role in the informal network have learned more, but instead argue that identifying the patterns of collaboration and participation provide a means to understand mechanisms for improving persistence and student success in physics.

## III. SOCIAL NETWORK ANALYSIS

SNA developed in the field of quantitative sociology. We have employed SNA as one tool for investigating the collaboration, participation, roles, and interactions among students who work together in the PLC. Wasserman and Faust's [17] text on SNA is the seminal work in this area. They identify four basic assumptions when conducting SNA:

(1) Actors and interactions are interdependent.
(2) Linkages allow flow (information, resources, etc.) between actors.
(3) Network models of individuals both constrain and provide opportunity for individual action.
(4) Network models conceptualize structures as representations of lasting patterns of relations among actors.

SNA is particularly useful for the analysis of relational data because the tools of SNA enable researchers to create visual representations of interactions and to quantify the interactions between actors in a network. Wasserman and Faust [17] should be referenced for greater insight into the various techniques available for SNA.

To facilitate the interpretation of our analysis we provide an overview of terms and measures and their interpretation within the domain of Social Network Analysis. First, the networks we consider are composed of *actors*, who are connected through relational *ties*, and the *attributes* of the actors. In our analyses, we use students as the actors as they are discrete social entities that participate in the network. Each actor has attributes: measurements of characteristics of the actors, which for us include data on the demographics as well as background information about the student. Ties, which are the linkages between actors and provide the structure to the network, in our study are represented by the identified collaborators for learning physics. For this study ties between students is the variable measured between all actors. SNA views ties as opportunities for exchanges, which in some cases may be information exchanges, monetary exchanges, or influence. In our study the ties can be seen as an opportunity to collaboratively build understanding or exchange ideas. Conversely, because of the relational nature of SNA, the absence of a tie between two actors constrains the interaction between actors, as the opportunity for exchange must go through some intermediary rather than passing directly between the two actors.

Consider the following example in which three students are trying to solve a physics problem: students A and B

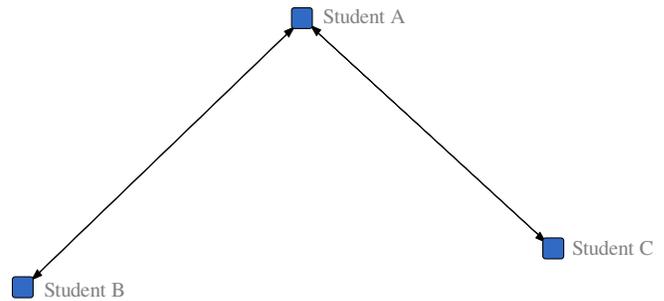

FIG. 1 (color online). Example sociogram of three students collaborating on homework.

work on homework together regularly, and student A also works with student C. This situation can be visualized using a sociogram, shown in Fig. 1. We see the three actors, student A, student B, and student C, as nodes on the sociogram. We also see ties between student B and student A and between student C and student A. However, there is an absence of a tie between student B and student C. Thus, if student B figures out a necessary step in solving the problem, a pathway exists for this step to be communicated directly with student A. However, a constraint for student C is that this new information must go through student A to reach student C.

Using these basic definitions, SNA has also established measures that can be used to characterize the structure of the relationships between actors within the network and the roles of the actors within this structure. For this paper, we consider two measures: *degree* and *geodesic distance*. These two measures are essential to calculating a third measure: *Bonacich's centrality* [18]. The degree of an actor within a network is defined as the total number of ties that a single actor is involved in. Degree is often viewed as a measure of the "activity" of the actor [17], p. 100. In terms of our study, a student who collaborates with a large number of students would have a high degree. In the sociogram in Fig. 1, student A has a larger degree than either student B or C because student A has two ties, while students B and C only have one tie each. Geodesic distance within a network is the minimum number of ties separating two actors.[1] Using the example of the three students depicted in Fig. 1, the geodesic distance between students A and B would be one and the geodesic distance between students B and C would be two, because there are two ties needed to create a path from student B to student C. In our study, a lower geodesic distance implies that there is both greater potential for flow of information between students and lesser constraint. These two measures, degree and geodesic distance, represent important features in our research: how active an actor is within a network and how well they are connected to other actors within the network.

---
[1]Incidentally this means the bar game "Six degrees of Kevin Bacon" would be better described as "Six Geodesic Distances of Kevin Bacon."

### Characterizing actors (students) and their roles in a network.

Typically, networks are analyzed either by looking at features of the network, such as how tightly the network is connected or by looking at the actors within the network. Our research will focus on the roles of the actors and how these roles reflect features of the structure of the Physics Learning Center. Because we know of no other research into characterizing the network of students in a semiformal learning environment, we are unable to compare with other networks. Instead, our goal in carrying out these analyses is to characterize the participation of actors in the PLC network. The usefulness of the tools of SNA has been demonstrated for this type of goal in a study from Dawson [19], who found significant positive correlations between students' degree and their score on Rovai's classroom community scale [20]. This study suggests SNA is useful for understanding the sense of academic and social integration that contributes to greater persistence. Our efforts at characterizing the participation in the PLC network will focus on the actors and their attributes and relating these to our overarching efforts to increase participation in physics of historically underrepresented students including women and minority students.

One of the opportunities afforded by network analysis is the ability to investigate the position an actor occupies within a network. Investigating how actors are embedded within a network can allow us to identify actors who are in favorable or unfavorable positions. Favorable positions include those who have more opportunities and fewer constraints; unfavorable positions face greater constraint and fewer opportunities. Investigating these positional advantages and disadvantages allows us to study power. Russell [21] defines power as the production of intended effects, but acknowledges that the manner of influence varies and that influence on opinion is one way to produce intended effects. According to Hanneman and Riddle [22], one of the important insights from network thinking is that power is inherently relational, meaning that the power one has within a network derives from the individual's relative position within the network. SNA is a quantitative approach to the relationships among actors, and therefore there exist tools for testing hypotheses related to positional advantage. In our analyses, we will use the attributes of actors to test if power is distributed according to any of the attributes of gender, ethnicity, or major.

Power is a difficult construct to measure, but SNA has established a measure of *centrality*, which is often interpreted as a proxy for power [18]. Centrality can be thought of as a measure of the advantage or disadvantage resulting from an actor's position within a network. This interpretation of centrality is consistent with the basic premises of SNA. If ties are seen as opportunities for exchange, then a person who has a large number of ties to other actors within a network has many opportunities to exchange or influence other actors within the network. Similarly, small geodesic distances between one actor and others means that the one actor has a short path to reach others and can relatively easily exchange information with other actors.

Centrality can be calculated in various ways; for our analyses we use Bonacich's approach to centrality [18]. Bonacich's approach constructs an eigenvector of centralities for each actor within a network, which accounts for both the centrality of the actor as well as the centrality of the actors with whom the actor is connected according to Eq. (1). In Eq. (1), $c_i$ represents the centrality of the $i$th actor, $\alpha$ is the standard centrality, $\beta$ symbolizes the extent to which an individual's status is a function of the statuses of those to whom he or she is connected, and $R$ represents a matrix of ties. The details and benefits of this approach for calculating centrality are detailed in Bonacich [18,23]. Bonacich's centrality is given by

$$c_i(\alpha, \beta) = \sum_j (\alpha + \beta c_j) R_{ij}. \quad (1)$$

Positional advantage, in our analyses, is an important construct because physics has historically been a predominantly male and White profession. In contrast, in the network we analyzed, the students are predominantly Hispanic, and the network has greater gender parity than in physics generally. Because we are interested in promoting equitable, inclusive models of science education, we use hierarchical regression analyses to construct models to uncover the variables that predict the centrality of students within the PLC informal learning environment. Thus we will be able to characterize participation in the PLC network and look for consistencies with our perspective on learning.

### IV. METHODS

Student users of the PLC completed an online survey that included background questions and asked, "What are the names of people that you work on homework with in the PLC?" One challenge of investigating social networks is the effort to be comprehensive about collecting data from all actors in the network. As a result, we sought responses from students enrolled in Modeling Instruction classes, members of the Society of Physics Students (which meets in the PLC), physics majors, and by posted signage in and around the PLC requesting that students complete the survey. These efforts generated 107 responses, of these, 99 students (from 7 different majors) completed the entire survey. These respondents included all levels of undergraduate students (1st year—6th year) and students with a variety of experience in physics (introductory classes to completing their degree). Six attributes of the actors collected on the self-reported survey included major, whether they had participated in a Modeling class,

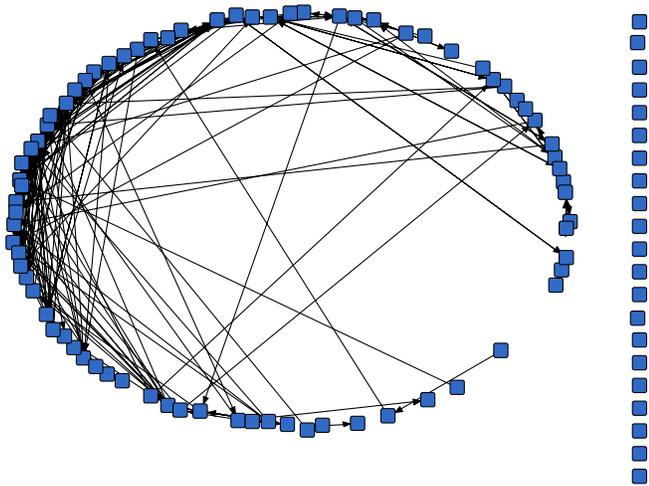

FIG. 2 (color online). Sociogram of all 99 students responding to the PLC survey. Each student is represented as a blue square and arrows represent ties between students. The line of 23 blue squares on the right-hand side of the diagram are students that had no tie to any other student in the network.

gender, ethnic background, number of days per week they spend in PLC, and time per week in PLC.

The 99 responses were then used to generate a 99 × 99 matrix of student interactions (1 means interaction present, 0 means not present). The demographic information was compiled into a 99 × 6 matrix, known as an attribute matrix in SNA. UCINET 6 [24], a standard SNA software package, and NETDRAW [25], a network visualization package integrated with UCINET, were used to complete all analyses. First, we constructed a sociogram including all 99 students (see Fig. 2).

Because we are not comparing this network to other networks, our analyses focus on the roles students play within the network. In order to do this, we first calculated the Bonacich centrality eigenvector. This eigenvector is made up of centrality values for each actor within the network: in this study a 99 × 1 vector. The values in this eigenvector can be thought of as the dependent variables for this study. Then we calculated correlation coefficients for each of the attributes (independent variables) with the dependent variable. Finally, using hierarchical multiple regression, we built a series of models which predict Bonacich centrality by regressing centrality on different attributes in the degree centrality eigenvector. The attributes used in this study include gender, ethnicity, major, time spent in the PLC, days per week in the PLC, and Modeling class participation.

Hierarchical multiple regression is a regression technique used to generate and compare the predictive models for a continuous dependent variable using different sets of independent variables. Typically, independent variables are entered in a specific order according to logical or theoretical considerations. In this study we entered variables in order of decreasing correlation with the centrality dependent variable. Table I lists the variables and order they were entered into our model. Table II shows the correlations among variables and correlations with the $\beta$ centrality for each variable. A new predictive model is created with each new independent variable entered. These models are then compared using a $z$ test for the differences between correlations as described by Tabachnick and Fidell [26]. In our case we constructed a model predicting centrality using only days per week in PLC as the predictor variable, then we constructed a second model using both days per week in PLC and major as predictor variables. We then used a $z$ test to determine if the additional variable improves the correlation with the observed data. The final predictive model was chosen as the model that significantly improves the prediction of centrality while using the least number of independent variables. This allows us to evaluate the importance that each independent variable plays in predicting centrality.

TABLE I. Independent variables used in hierarchical multiple regression, variable type, and order of entry into regression model.

| Order | Independent variable | Variable type (possible values) |
|---|---|---|
| 1 | Days per week in PLC | Ordinal (1–2, 3–4, 5, 6–7) |
| 2 | Major | Categorical (Physics 1; Other 2) |
| 3 | Time per week in PLC (in hours) | Ordinal (1–2, 3–4, 5–6, 7–8, 8+) |
| 4 | Ethnicity | Categorical (Majority 1; Underrepresented 2) |
| 5 | Introductory course type | Categorical (Modeling 1; Traditional 2) |
| 6 | Gender | Categorical (Female 1; Male 2) |

TABLE II. Correlation matrix of independent variables used in hierarchical multiple regression.

|  | Days per week | Major | Time per week | Ethnicity | Introductory course type | Gender | Correlation with $\beta$ centrality |
|---|---|---|---|---|---|---|---|
| Days per week | 1.00 | 0.228 | 0.446 | −0.033 | −0.032 | 0.038 | 0.549 |
| Major | ⋯ | 1.00 | 0.238 | −0.196 | −0.145 | 0.165 | 0.540 |
| Time per week | ⋯ | ⋯ | 1.00 | 0.064 | 0.130 | −0.118 | 0.313 |
| Ethnicity | ⋯ | ⋯ | ⋯ | 1.00 | −0.003 | −0.181 | −0.166 |
| Introductory course type | ⋯ | ⋯ | ⋯ | ⋯ | 1.00 | −0.034 | −0.160 |
| Gender | ⋯ | ⋯ | ⋯ | ⋯ | ⋯ | 1.00 | 0.128 |

Within Social Network Analysis the basic assumption is that actors within a network are interdependent. This interdependency means that standard parametric statistics are not appropriate due to the violation of the standard parametric assumption of independence of measures. Instead, statistical analyses in SNA rely on bootstrap methods and Monte Carlo simulations; therefore, all regression analyses are carried out in UCINET 6.0. However, while the regression analysis is carried out using UCINET, the interpretation of results is consistent with standard regression.

## V. RESULTS AND ANALYSIS

The results of the hierarchical multiple regression, summarized in Table III, indicate that the model which best predicts centrality in the Physics Learning Center, with the fewest number of variables, includes both number of days per week spent in PLC and major variables. In comparing the models, we found that after entering days per week in PLC in the equation, the model significantly predicted the observed data (see model 1 in Table III), $R^2 = 0.301$, $F(1, 106) = 41.81$, $p < 0.001$. The model that added major into the equation (see model 2 in Table III) also significantly predicted the observed data: $R^2 = 0.482$, $F(2, 108) = 44.74$, $p < 0.001$. Further,

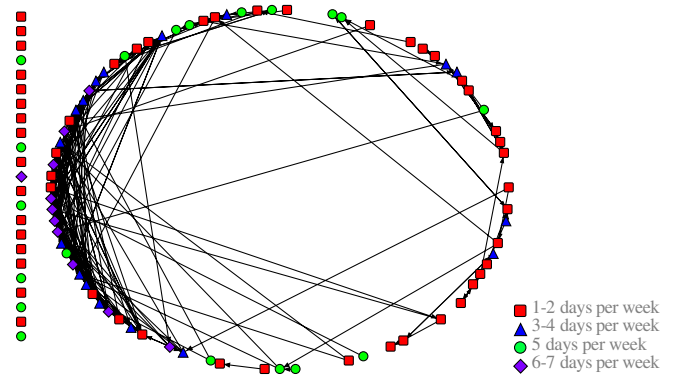

FIG. 3 (color online). Sociogram of Physics Learning Center participation. Nodes represent students, ties represent self-reported collaboration, red squares spent 1–2 days per week, blue triangles spent 3–4 days per week, green circles spent 5 days per week, and purple diamonds spent 6–7 days per week in the Physics Learning Center.

adding major to the model already including days per week in PLC resulted in a significant improvement in the correlation between the model and the observed centrality data (a 2.11 standard deviation improvement in correlation). While model 3, which added time per week in PLC to the model, significantly predicted the correlation data with $R^2 = 0.483$, $F(3, 108) = 29.53$, $p < 0.001$, a $z$ test indicated that including time per week in PLC did not significantly improve the correlation over model 2. Including the time per week in the PLC only improved the correlation by 0.03 standard deviations, so it does not greatly improve our predictive model. Similarly, other variables when added to the equation (see models 4–6 in Table III) did not significantly improve the correlation between model 2 and the observed data. This pattern indicates that the other variables (gender, ethnicity, introductory course type) do not improve prediction of centrality over the days per week in PLC and major variables. To represent the results of this regression analysis, we created sociograms of the ties between actors grouped by these different attributes (see Figs. 3 and 4).

TABLE III. Comparison of hierarchical multiple regression models. (CI is confidence interval.)

| | Variables included in the model | $R^2$—Correlation with centrality [95% CI] | $\Delta$ standard deviations in model fit |
| --- | --- | --- | --- |
| Model 1 | Days per week | 0.301 [0.155–0.453] | ... |
| Model 2 | Days per week, major | 0.482 [0.331–0.615] | 2.11[a] |
| Model 3 | Days per week, major, time per week | 0.483 [0.333–0.616] | 0.03 |
| Model 4 | Days per week, major, time per week, ethnicity | 0.487 [0.337–0.619] | 0.13 |
| Model 5 | Days per week, major, time per week, ethnicity, introductory course type | 0.496 [0.347–0.627] | 0.34 |
| Model 6 | Days per week, major, time per week, ethnicity, introductory course type, gender | 0.497 [0.348–0.628] | 0.06 |

[a]Indicates significant difference in the correlation between the current model and previous model.

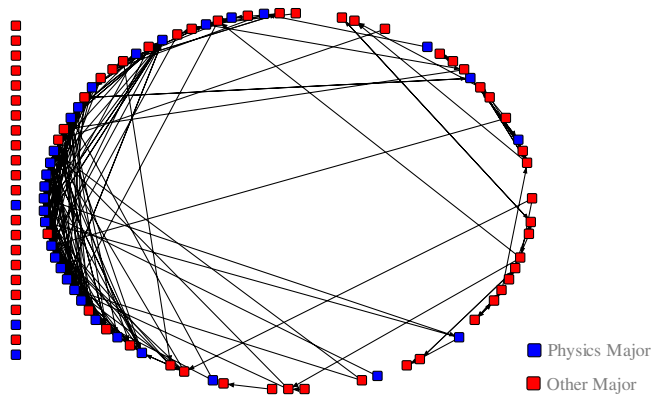

FIG. 4 (color online). Sociogram of Physics Learning Center participation. Nodes represent students, ties represent self-reported collaboration, red squares are not physics majors, blue squares are physics majors.

Interpreting these results suggests several important conclusions about participation in the student learning community centered in the Physics Learning Center. The first is that two predictor variables (major and days per week) allow us to predict centrality, meaning that students have a degree of control over their centrality in the PLC. Unsurprisingly, visiting the PLC regularly increases the likelihood that a student is a central participant within the learning community. The sociogram in Fig. 3 depicts this finding as the students represented with purple diamonds (6–7 days per week) tend to be farther on the left-hand side of the circle and tend to have a greater density of connections with other students. Students who report visiting the PLC 1–2 days per week are represented with red squares, many of whom are isolates (students not connected with any other students in a line on the left-hand side of the diagram) or farther to the right of the diagram where connections are more sparse indicating lesser centrality. The other sociogram, Fig. 4, shows physics majors more concentrated in the heavily connected portion of the diagram, which is consistent with results from model 2 where the major variable improved the prediction of the centrality of students. Both these variables, days per week in the PLC and major, are variables that can be controlled by the student to an extent. These results indicate that participation and centrality in the PLC is something that can be achieved by any student, assuming the lack of any latent variables not measured as a part of this study. Conversely, variables that are out of the control of the students, such as gender and ethnicity, do not add to the prediction of centrality in the PLC. The absence of predictive power from variables such as gender and ethnicity indicates that participation in the learning community is equitably distributed. In other words, students do not appear to be excluded due to gender or ethnic background.

## VI. DISCUSSION AND CONCLUSIONS

The multiple regression analysis of Bonacich centrality suggests that membership in the PLC learning community is equitably distributed in terms of student participation and students' centrality. Gender and ethnicity variables did not significantly improve the prediction of student centrality, indicating that the learning community is not exclusive of students of a particular gender or ethnicity. This finding contrasts with other classroom-based measures, specifically with gender learning gaps in introductory physics [10,27,28]. Further, these findings are consistent with sociocultural and participatory views on learning [15,29–31] and provide plausible suggestions for efforts which could support promotion of greater gender and ethnic diversity among physics majors. It should be noted that the results presented here are from FIU, which is a Hispanic-Serving Institution with a majority minority population, and Hispanic participation in learning communities has been shown to be of importance in other studies [2]. These findings suggest that in order to promote the retention and persistence of all students, physics departments could take active steps to provide pathways and access to participation in a learning community. Taking these steps, however, requires departments to attend to a complex set of interrelated features, many of which are difficult to directly measure and may not be tied to any specific class. This underlying complexity of the educational process has been described by Forsman et al. [32] who have proposed using complexity thinking as a means by which to understand student persistence. We expect that further work on complexity theory and the role of SNA in complexity thinking could prove productive for understanding issues of persistence in physics.

A second conclusion is that social network analysis holds significant promise for the description and analysis of student learning communities and therefore has potential impact on methods of supporting students' participation, retention, and persistence in physics. SNA is a useful approach to investigating relational variables: including collaboration such as we have investigated, but also variables which could easily include interactions between and among students, teachers, learning assistants, concepts, models, and many others. Though it can be used as a research methodology, it also provides rationales for practical decision making, such as reason for a department to support the existence of a multipurpose space such as the Physics Learning Center at FIU. The extent of the network and relationships within that network would have been nearly impossible without the Physics Learning Center space. Fostering interactions such as those that take place in the PLC and supporting students through academic and social integration are goals that are reasonable for physics departments to consider. Providing space for the network to develop is one reasonable approach to enhancing collaboration which could reasonably contribute to increased diversity in physics. Identifying this without the tools available for SNA would have been nearly impossible.

We have illustrated how SNA provides a rich set of methodological tools for describing learning communities and identifying features inaccessible through classroom measures alone. Future directions include using complexity thinking as a way of modeling the efficacy of the system [32] and to look for other advantages afforded by participation in the learning community. Further, we anticipate that SNA can prove useful in describing other departments that are successfully retaining students.

## ACKNOWLEDGMENTS

The authors would like to thank Aaron Warren for his valuable contributions to the analysis and feedback on the use of SNA. This work is supported by NSF Grant No. PHY-0802184.